\documentstyle[11pt]{article}
\begin{document}
\renewcommand{\theequation}{\arabic{section}.\arabic{equation}}
\parskip = 0 pt
\baselineskip = 16 pt
\parindent = 20 pt
\pagenumbering{roman}
\centerline{\LARGE Complete positivity and subdynamics}\par
\centerline{\LARGE in quantum field theory}\par
\vskip 15  pt
\centerline{B.~Vacchini\footnote{Dipartimento di Fisica
dell'Universit\`a di Milano and Istituto Nazionale di Fisica
Nucleare, Sezione di Milano, Via Celoria 16, I-20133, Milan,
Italy. E-mail: bassano.vacchini@mi.infn.it}}
\vskip 15 pt
\centerline{\sc Abstract}\par
\vskip 15 pt
{
\baselineskip=12pt
The relevance that the property of complete positivity has
had in the determination of quantum structures is briefly
reviewed, together with recent applications to neutron
optics and quantum Brownian motion.
A possible useful application and generalization of this property
to the description of macroscopic systems inside  quantum
mechanics is discussed on the basis of recent work on
the derivation of  subdynamics in Heisenberg picture
of slowly varying degrees of freedom inside nonrelativistic
quantum field theory.
}\par
\vskip 25 pt
\hrule
\par
\vfill\eject
\par
\section{INTRODUCTION}
\pagenumbering{arabic}
\par
Even though a thorough understanding of quantum mechanics
(QM) is still far away and many different readings and
interpretations of QM coexist in the scientific community,
major progress has been made in the study and determination
of quantum structures, both from a logical and mathematical
point of view. The logical studies on QM originated in the
seminal paper of Birkhoff and von Neumann (1936) and
have by now reached important results, mainly thanks to the
basic notions of effect and effect algebras (for a recent
review see Dalla Chiara and Giuntini, n.d.).
On the physical and mathematical side it is hardly feasible
to do justice to the full inventory of mathematical tools
and properties that have been introduced and understood to
be relevant to the realm of QM, and especially  quantum
measurement theory. In fact all studies concerning  quantum
structures aim at a better understanding of the foundations
of QM and therefore often address either directly or
indirectly the subject of measurement theory.
\par
As far as this paper is concerned we are mainly interested
in notions and tools that have grown out of the so called
operational approach to QM (for a recent review see
Busch {\it et al.}, 1995), whose
physical origin is to be traced back to the original work of
Ludwig and coworkers on the foundations of QM (Ludwig,
1983). As a matter of fact the work of Ludwig, which was originally
intended to obtain a reconstruction of the Hilbert space
structure of QM based on a  statistical formulation of the
theory relying on a classical, objective  description of
preparation and  registration  apparatuses, apart from
giving a fundamental contribution to the foundations of QM,
has also somehow ``incidentally'' led to the introduction of
concepts, such as those of effect and operation, that are by
now very useful in applications of QM, such as for example
quantum optics and quantum information. In particular the
notions of effect and operation, together with more general
and refined mathematical tools derived from these, such as
those of coexistent observable, effect valued measure and
instrument, have made the very formulation of continuous
measurement feasible (Srinivas and Davies, 1981; Barchielli {\em et
al.}, 1983),
perhaps one of the major achievements within
quantum measurement theory. This result, though somehow to
be expected on the basis of experimental evidence, think for
example of the shelving effect (Dehmelt, 1990), was certainly
not obvious in the early days of QM, and provides some
evidence substantiating the conjecture of Ludwig about the
possibility of founding QM on macroscopic systems to be
objectively described in a suitable trajectory space.
\par
In the sequel we will deal in particular with the notion
of complete positivity (CP), whose relevance to the realm of
QM was first realized by Kraus (1971) and Lindblad (1976).
This property has recently played an
outstanding  role in the study of open  quantum  systems,
both at a fundamental as well as at a phenomenological level
and we will give a few important examples in which it has
actually led to the determination of specific quantum
structures. We will then argue how CP, and in particular
generalized, mathematically less stringent version of this
property, might play a role in the determination of
subdynamics inside nonrelativistic  quantum field theory.
This will be done on the basis of recent work carried out in
the fields of neutron optics (Lanz and Vacchini, 1997a,b) and quantum Brownian
motion (Vacchini, 2000),
as well as a recently outlined approach for
the description of the dynamics of slowly varying degrees of
freedom within a macroscopic system (Lanz {\it et al.}, 1997;
Lanz and Vacchini, 1998). This should
shed some light on possible useful extensions of the
property of CP from one-particle QM to the realm of quantum
field theory applied to many-body  systems.
\par
\section{COMPLETE POSITIVITY}
\par
Let us now briefly introduce the definition of CP.
The most general representation of the preparation
of a physical system
described in a Hilbert space ${\cal H}$ is given by
a statistical operator, that is to say
an operator in the space
${\cal TC}({\cal H})$
of trace class operators on ${\cal H}$, positive and with
trace equal to one.
In particular we call ${\cal K}({\cal H})$ the convex set of
statistical operators
        \[
        {\cal K}({\cal H})=
        \left \{
        {\hat \varrho}\in {\cal TC}({\cal H}) | \,
        {\hat \varrho} \geq 0, \> {\rm Tr}{\hat
        \varrho}=1
        \right \}
        \]
Consider now a mapping ${\cal U}$ defined on the space of
trace class operators into itself
        \[
        {\cal U}:
        {\cal TC}({\cal H}) \longrightarrow
        {\cal TC}({\cal H})
        \]
possibly corresponding to a
Schr\"odinger picture description on the states.
We say that the map ${\cal U}$ is completely positive, or
equivalently has the property of complete positivity (Kraus,
1983; Alicki and Lendi, 1987), if and only if the adjoint map ${\cal U}'$
acting on the space ${\cal B}({\cal H})$ of bounded  linear
operators, dual to ${\cal TC}({\cal H})$,
        \[
        {\cal U}':
        {\cal B}({\cal H}) \longrightarrow
        {\cal B}({\cal H})
        \]
and therefore corresponding to an Heisenberg picture description in terms of
observables, satisfies the inequality
        \begin{equation}
        \label{CP}
        \sum_{i,j=1}^n
        \langle\psi_i\vert
        {\cal U}'
        \left(
        {\hat{B}}{}_i^{\scriptscriptstyle\dagger}
        {\hat {B}}{}_j
        \right)
        \vert\psi_j\rangle
        \geq 0
        \qquad
        \forall n\in{\bf N}, \quad
        \forall
        \left \{
        \psi_i
        \right \}
        \in {\cal H}
        , \quad
        \forall
        \{
        {\hat B}_i
        \}
        \in {\cal B}({\cal H})
        \end{equation}
For $n=1$ one recovers the usual notion of positivity, while
for bigger $n$ this is actually a nontrivial requirement.
\par
It is immediately seen that if ${\cal U}'$ has the
factorization property
        \[
        {\cal U}'
        (
        {\hat{A}}^{\scriptscriptstyle\dagger}
        {\hat {B}}
        )
        =
        \left[
        {\cal U}'
        (
        {\hat{A}}
        )
        \right]^{\scriptscriptstyle\dagger}
        {\cal U}'
        (
        {\hat {B}}
        )
        \qquad
        \forall
        {\hat A}, {\hat B}
        \in {\cal B}({\cal H})
        \]
then it is CP, so that any unitary
evolution is CP. In this sense one can see CP as a property
that is worth retaining when shifting from the unitary
dynamics for closed systems to a more general dynamics for
the  description of open systems. In fact the general
physical argument for the introduction of CP is the
following.
Consider a system ${\cal S}_1$ described in ${\cal H}_1$,
whose dynamics is given by the family of mappings
        \[
        {\cal U}:
        {\cal TC}({\cal H}_1) \longrightarrow
        {\cal TC}({\cal H}_1)
        \]
and an $n$-level  system ${\cal S}_2$ described in ${\cal
H}_2 = {\bf C}^n$, whose dynamics can be neglected, so that
${\hat H}_2=0$. Because the two  systems do not interact,
the map $\tilde{{\cal U}}$ describing their joint evolution
        \[
        \tilde{{\cal U}}:
        {\cal TC}({\cal H}_1 \otimes {\bf C}^n) \longrightarrow
        {\cal TC}({\cal H}_1 \otimes {\bf C}^n)
        \]
will be simply given by the tensor product
$\tilde{{\cal U}} = {\cal U} \otimes {\bf 1}$. But the
dynamical map  $\tilde{{\cal U}}$ must of course be positive
and this is equivalent to the requirement that ${\cal U}$ be CP.
\par
The property of CP has already shown to be particularly relevant in the
determination of quantum structures, and in the sequel we
will give two important examples in this connection.
\hfill \break
Let us consider first the notion of operation,
which is the basic tool for the  description of changes
experienced by a physical  system. An operation ${\cal F}$
is a positive linear map
acting on the space of trace class operators and sending
statistical operators in positive operators with trace less
or equal than one
        \[
        {\cal F}:
        {\cal TC}({\cal H}) \longrightarrow
        {\cal TC}({\cal H})
        \qquad
        0 \leq {\rm Tr} {\cal F}({\hat \varrho}) \leq 1
        \qquad
        \forall {\hat \varrho}\in {\cal K}({\cal H})
        \]
Operations describe the repreparations of a statistical
collection based on some measurement outcome
and the connection between such mappings and CP
was first considered by Kraus (1983) and Hellwig (1995).
The requirement of CP, according to the Stinespring
representation theorem, determines the general structure
of such mappings to be:
        \[
        {\cal F}({\hat T})
        =
        \sum_{k\in K}
        {\hat {A}}{}_k^{\scriptscriptstyle\dagger}
        {\hat T}
        {\hat {A}}{}_k
        \quad
        \forall\, {\hat T}\in {\cal TC}({\cal H})
        \qquad
        K\subset {\bf N},
        \quad
        0 \leq
        \sum_{k\in K}
        {\hat {A}}{}_k^{\scriptscriptstyle\dagger}
        {\hat {A}}{}_k
        =
        {\hat F}
        \leq {\bf 1}
        \]
where ${\hat F}$ is the effect associated to the operation,
even though not uniquely specifying it. Let us note that the
notion of operation, previously considered only in the
studies of fundamental nature about QM and quantum
measurement theory, is now being used by a much broader
physical community thanks to the applications in quantum
optics and more recently  quantum communication and  quantum
information theory. For example CP trace preserving
operations are by now taken as the standard definition of
quantum channels (Schumacher, 1996).
\par
The other example comes from the field of quantum
dynamical semigroups (Alicki and Lendi, 1987), used for the description of
the irreversible dynamics of open quantum
systems, typically the reduced dynamics of systems
interacting with an external system, such as a heat bath or
a measuring instrument.
In Heisenberg picture quantum dynamical semigroups
are given by collections of positive mappings
        \[
        {\cal U}'_t:
        {\cal B}({\cal H}) \longrightarrow
        {\cal B}({\cal H})
        \qquad t\geq 0,
        \qquad
        {\cal U}'_0 {\bf 1}={\bf 1}
        \]
which satisfy the
semigroup composition property
        \[
        {\cal U}'_s
        {\cal U}'_t
        =
        {\cal U}'_{s+t}
        \qquad s,t\geq 0
        \]
and which are normal.        
Under these conditions a generally unbounded generator ${\cal
L}'$ defined on an ultraweakly dense domain exists such that
        \[
        {
        d
        \over
         dt
        }
        {\cal U}'_t {\hat B}
        =
        {\cal L}'{\cal U}'_t {\hat B}
        \]
for all ${\hat B}$ in the domain. If one further asks
the semigroup to be norm continuous, so that the generator
is a
bounded map, it can be shown, as has been done by Lindblad
(1976) in a
by now very famous paper, that CP determines the
general expression for the
generator to be of the form
        \[
        {\cal L}' {\hat B}
        =
        {i \over \hbar}
        [
        {\hat {{H}}},
        {\hat B}
        ]
        -{1\over 2}
        \left \{  
        \sum^{}_j
        {\hat V}{}_j
        {\hat V}{}_j^{\scriptscriptstyle \dagger}
        , {\hat B}
        \right \}
        +  
        \sum^{}_j
        {\hat V}{}_j^{\scriptscriptstyle
        \dagger}
        {\hat B}
        {\hat V}{}_j
        \]
        \[
        {\hat V}{}_j,
        \sum^{}_j
        {\hat V}{}_j
        {\hat V}{}_j^{\scriptscriptstyle \dagger}
        \in {\cal B}({\cal H}),
        \qquad
        {\hat H}{}_j =
        {\hat H}{}_j^{\scriptscriptstyle \dagger}
        \in {\cal B}({\cal H})
        \]
This structure of master equation, possibly allowing for
unbounded  operators or even quantum fields,
appears in many applications
in very different fields of physics and is often taken as
starting point for phenomenological approaches.
It accounts for a non-Hamiltonian dynamics and has
been extensively used in the
formulation of continuous measurement theory and especially
in quantum optics.
\par
\section{SUBDYNAMICS IN QUANTUM FIELD THEORY}
\par
Inspired by Ludwig's work we take the attitude according to
which any macroscopic physical system is actually defined by
the preparation  procedure which separates it from the rest
of the world. Such systems are to be described in terms of
interacting and confined quantum fields, whose choice
depends on the considered description level. In fact
realistic confinement and isolation can only be considered
with reference to a coarse graining of the time scale, which
allows to consider a break-up of the correlations with the
environment and to replace the actual physical walls by
suitably idealized boundary conditions.
For the description of the system on the given time scale we
look for the subdynamics of a subset of variables, which are slowly
varying on this time scale.
Contrary to older attempts at the derivation of a master
equation for the reduced dynamics of the  statistical
operator, we work in Heisenberg picture on
a restricted set of observables, whose choice depends on the
particular features of the  system and of the preparation
procedure. In considering the subdynamics of these
quantities, on the basis of the experienced gained with
one-particle QM, we will ask for the property of CP, a
viewpoint also shared by Streater (1995).
Using the field theoretical
formalism of second quantization observables have the
general expression
        \[
        {\hat A}
        =
        \sum_{h_1\ldots h_n \atop k_1\ldots k_n}
        a^{\scriptscriptstyle\dagger}_{h_n} \ldots
        a^{\scriptscriptstyle\dagger}_{h_1}
        {\cal A}^{\phantom{\scriptscriptstyle\dagger}}
        (h_n,\ldots,h_1,k_1,\ldots,k_2)
        a_{k_1} \ldots a_{k_n}
        \]
so that, recalling (\ref{CP}), this typical structure in
terms of a block of annihilation operators and a block of
creation operators is a natural candidate for the
requirement of CP. Moreover typical slow variables will
be positive densities of conserved quantities, such as mass
and energy, which are easily expressed in a field formalism
and whose positivity has to be preserved throughout the
evolution.
\par
Before going over to consider some concrete examples of this
formal scheme, let us briefly mention the possible relevance
of this approach to the foundations of QM. A general
formulation of the theory of macroscopic systems in terms of a
non-Hamiltonian irreversible dynamics for a selected set of
observables could be the starting point for their
description inside continuous measurement theory, thus
possibly recovering a classical, objective description,
but objectively described  macroscopic systems build the
basis on which  Ludwig founded his remarkable approach to
the foundations of QM.
\par
The simplest application of the proposed formal scheme
consists in the description of a microsystem interacting
with a macroscopic system supposed to be at equilibrium,
typically a particle interacting with matter. In this case
the particle constitutes the slow degree of freedom with
respect to the fast relaxation time of the macroscopic
system and a Markov approximation for the particle's
dynamics should hold, provided the two time scales are
separated. We here briefly sketch the main points in
formalism and calculation (for details see Lanz and
Vacchini, 1997a). In second
quantization the Hamiltonian reads
        \[  
        {\hat H}={\hat H}_0 + {\hat H}_{\rm{m}} + {\hat V}
        \qquad  
        \qquad  
        {\hat H}_0 = \sum_f
        {E_f} {a^{\scriptscriptstyle \dagger}_{f}} {a_{{f}}}  
         \qquad  
         \qquad  
        {\left[{{a_{{f}}},{a^{\scriptscriptstyle 
        \dagger}_{g}}}\right]}_{\mp}=\delta_{fg} 
        \]  
The whole  system is described in ${\cal H}$, while the
Hilbert space for the  microsystem ${{\cal H}^{(1)}}$ is
spanned by the energy eigenstates $u_f$, so that
${a_{{f}}}$ is the destruction operator for the microsystem,
whose statistics is left unprejudiced, in
the state $u_f$. ${\hat H}_{\rm{m}}$ describes matter and
${\hat V}$
is the interaction potential.
Being interested in the description of a single particle, we
take a statistical operator of the form
        \[  
        {\hat \varrho}=
        \sum_{{g} {f}}{}  
        {a^{\scriptscriptstyle \dagger}_{g}}  
        {{\hat \varrho}_{\rm{m}}} {a_{{f}}}
        {{\varrho}}^{(1)}_{gf}    ,
        \]  
where ${{\hat \varrho}_{\rm{m}}}$ describes matter and
${{\varrho}}^{(1)}_{gf}$ is a positive matrix with trace equal to
one.
In order to extract the subdynamics of the particle we
consider field observables of the form
${{\hat A}}
        = \sum_{h,k}
        {a^{\scriptscriptstyle \dagger}_{h}}
        {A}_{hk}^{(1)}
        {a_{k}}
$
and exploit the reduction formula
        \[
        {\hbox{\rm Tr}}_{{\cal H}}
        \left(  
        {{\hat A}{\hat \varrho}}
        \right)
        =
        {\hbox{\rm Tr}}_{{{\cal H}^{(1)}}}
        \left(  
        {{\hat {\sf A}}^{(1)} {\hat \varrho}^{(1)}}
        \right)
        \]  
where ${(1)}$ denotes  operators in  ${{\cal H}^{(1)}}$.
What we have to evaluate is the time evolution, in
Heisenberg  picture, of bilinear structures of field
operators ${{a^{\scriptscriptstyle \dagger}_{h}}{a_{k}}}$,
on a suitable time scale, much longer than the microphysical
interaction time $\tau_0$. To do this we exploit a
superoperator formalism, thus considering maps acting on the
algebra of creation and destruction  operators, for example
        \[
        {\cal H}^{'}_0 ({{a^{\scriptscriptstyle
        \dagger}_{h}}{a_{k}}})
        ={i \over \hbar} [{\hat H}_0 + {\hat H}_{\rm{m}},
        {{a^{\scriptscriptstyle \dagger}_{h}}{a_{k}}}]
        =
        {
        i
        \over
         \hbar
        }
        (E_h - E_k) {{a^{\scriptscriptstyle \dagger}_{h}}{a_{k}}}
        \]
and exploit techniques of scattering theory. We thus work in
a Markov approximation considering slow variables on the
given time scale, so that the quasi-diagonality condition
$
        \hbar
        /
        |
        E_h - E_k  
        |
        \geq
        t
        \gg
        \tau_0
$
should be satisfied. We also suppose suitable smoothness
properties of the T-matrix, so that there are no bound
states. As a result we obtain the following structure for
the evolution mapping on a time $t$ which is small with
respect to the particle's dynamics, though much larger than
the relaxation time of the  macrosystem
        \[
        {{\cal U}'(t)}
        \left(  
        {{{ a}^{\scriptscriptstyle \dagger}_{h}}{{ a}_{k}}}
        \right)  
        =
        {{{ a}^{\scriptscriptstyle \dagger}_{h}}{{ a}_{k}}}
        +
        t {\cal L}'
        {{{ a}^{\scriptscriptstyle \dagger}_{h}}{{ a}_{k}}}
        \]
where the generator restricted to this typical bilinear
structure of field  operators in the quasi-diagonal case is
given by:
        \[
        {\cal L}'
        \left(  
        {{{ a}^{\scriptscriptstyle \dagger}_{h}}{{ a}_{k}}}
        \right)  
        =
        {i\over\hbar}  
        \left[  
        {\hat H}_0 + {\hat V}^{[1]},
        { a}^{\scriptscriptstyle \dagger}_{h}
        { a}_{k}
        \right]  
        - {1\over \hbar}  
        \left\{
        \left[  
        {\hat \Gamma}^{[1]} , { a}^{\scriptscriptstyle\dagger}_h
        \right]  
        { a}_k
        -  
        { a}^{\scriptscriptstyle\dagger}_h
        \left[
        {\hat \Gamma}^{[1]}, { a}_k
        \right]  
        \right\}
        +  
        {1\over\hbar} \sum_\lambda  
        {\hat R}^{[1]}_{h \lambda}{}^{\dagger}
        {\hat R}^{[1]}_{k \lambda}
        \]
the index ${[1]}$ denoting one-particle operators,
${\hat V}^{[1]}$ and ${\hat \Gamma}^{[1]}$ being linked respectively to
the
self-adjoint and anti-self-adjoint part of the T-matrix. Let
us note that due to the presence of the minus sign the term
between curly brackets cannot be rewritten as a simple
commutator. CP of the mapping ${\cal U}' (t)$ restricted to
these simple bilinear field structures
        \[
        \sum_{i,j=1}^n
        \langle\psi_i\vert
        {\cal U}'(t)
        \left(
        \sum_{hk}
        a^{\scriptscriptstyle\dagger}_h
        \langle h\vert {\hat{\sf
        B}}{}_i^{\scriptscriptstyle\dagger}
        {\hat {\sf B}}{}_j \vert k \rangle    a_k
        \right)
        \vert\psi_j\rangle
        \geq 0
        \]
can be seen from the decomposition which holds true for
an infinitesimal positive time $dt$
        \begin{eqnarray*}
        a^{\scriptscriptstyle\dagger}_h a_k + dt    \,
        {\cal L}'
        (
        a^{\scriptscriptstyle\dagger}_h a_k
        )
        &=&
        {
        \left \{
        {
        a_h+\frac{i}{\hbar}
        dt
        \left[
        {\hat H}_0 + {\hat V}^{[1]}
        ,a_h
        \right]
        -\frac{dt}{\hbar}
        \left[
        {\hat \Gamma}^{[1]} , { a}_h
        \right]  
        }
        \right \}
        }^{\dagger}
        \\
        &&
        \hphantom{=}
        \times
        \left\{
        {
        a_k+\frac{i}{\hbar}
        dt
        \left[
        {
        {\hat H}_0 + {\hat V}^{[1]}
        ,a_k
        }
        \right]
        -\frac{dt}{\hbar}
        \left[  
        {
        {\hat \Gamma}^{[1]} , { a}_k
        }
        \right]
        }
        \right\}
        {}+ \frac{dt}{\hbar} \sum_{\lambda}
        {\hat R}^{[1]}_{h \lambda}{}^{\dagger}
        {\hat R}^{[1]}_{k \lambda}
        \end{eqnarray*}
One can also check that particle number conservations holds,
so that
${\cal L}'({\hat N})=0$, where ${{\hat N}}=\sum_f {{a^{\scriptscriptstyle
        \dagger}_{f}}{a_{f}}}$.
In order to consider the microsystem's degrees of freedom,
we take a partial trace over matter
        \[
        {
        d 
        \over
        dt
        } {\varrho}^{(1)}_{kh}
        =
        {\hbox{\rm Tr}}_{{{{\cal H}}}}
        \left(
        {\cal L}'
        \left(
        a^{\scriptscriptstyle\dagger}_h a_k 
        \right)
        {{\hat \varrho}_{\rm{m}}}
        \right)
        \]
and obtain the following master equation of the Lindblad
form for the subdynamics of the particle, thus automatically
ensuring CP
        \[
        {
        d 
        \over  
                      dt
        } {\hat {\sf \varrho}}^{(1)}
        =  
        -{i \over \hbar}  
        \left[{\hat {{\sf H}}}_0^{(1)}
        +  
        {\hat {\sf V}}^{(1)},
        {\hat {\sf \varrho}}^{(1)}
        \right]
        -{1\over\hbar}  
        \left \{  
        {  
        {\hat {\sf \Gamma}}^{(1)}
        , {\hat {\sf \varrho}}^{(1)}
        }
        \right \}
        +  
        {1 \over \hbar}  
        \sum^{}_{{\xi\lambda  }}
        {\hat {\sf L}}_{\lambda\xi}^{(1)} {\hat \varrho}^{(1)}
        {{\hat {\sf L}}{}_{\lambda\xi}^{(1)}{}^{\scriptscriptstyle \dagger}}\ ,
        \]
where ${\hat {\sf V}}^{(1)}$ and ${\hat {\sf \Gamma}}^{(1)}$
are still  linked to the
self-adjoint and anti-self-adjoint part of the T-matrix
averaged over matter, and particle number conservation
implies
$
        {\hat {\sf \Gamma}}^{(1)}
        =
        1/2
        \sum^{}_{{\xi\lambda  }}
        {{\hat {\sf L}}{}_{\lambda\xi}^{(1)}{}^{\scriptscriptstyle \dagger}}
        {\hat {\sf L}}{}_{\lambda\xi}^{(1)}
$.
This master equation is well suited to describe both
coherent and incoherent behavior. Apart from a commutator
term analogous to a Liouville - von Neumann equation, it has
an anticommutator term, which might be obtained by
introducing a complex potential in the Schr\"odinger
equation and a mixture term which is only characteristic of the
formalism of the  statistical operator. It
will here be applied to two examples.
\par
In the first case we consider the mainly coherent
interaction of thermal neutrons with homogeneous samples,
the so called neutron optics, relevant for the description
of neutron interferometry experiments (for references and
detail see Lanz and Vacchini, 1997b).
The phenomenological Ansatz used for the description of
neutron matter interaction is the Fermi pseudopotential
        \[
        {\hat T}=
        {  
        2\pi \hbar^2  
        \over  
        m  
        }  
        b
        {\int d^3 \! {\bf{r}} \,}
        {\psi}^{\scriptscriptstyle\dagger} ({\bf{r}})
        \delta^3 ({\hat {{\sf x}}}-{{\bf{r}}})
        {\psi} ({\bf{r}})
        \]
a local potential parameterized by the coherent scattering
length $b$. Using this Ansatz and leaving out the
incoherent, dissipative terms we obtain the usual wavelike
description of the interaction, in terms of a refractive
index
$
n \simeq
[
        1- (\lambda^2 / 2\pi) b n_{\rm{o}}
]
$
depending on the density $n_{\rm{o}}$ of particles in the medium.
The dissipative  contributions instead can be expressed in terms of
the dynamic structure function of the medium
        \[
        S ({\bf{q}},\omega)=
        {  
        1  
        \over  
         2\pi N  
        }  
        \int dt \,  
        {\int d^3 \! {\bf{x}} \,}
        e^{  
        -i(\omega t - {\bf{q}}\cdot{\bf{x}})
        }  
        {\int d^3 \! {\bf{y}} \,}
        \left \langle  
        {\hat N}({\bf{y}})
        {\hat N}({\bf{x}}+{\bf{y}},t)
        \right \rangle  
        \]
where $\hbar\omega$ and $\hbar{\bf{q}}$ denote energy and
momentum transfer, and ${\hat N}({\bf{x}})=
        {\psi}^{\scriptscriptstyle\dagger} ({\bf{x}})
        {\psi} ({\bf{x}})
$  local densities. These terms lead to an imaginary
correction to the optical potential
        \[
        {\hat {{\sf U}}}  =  
        {  
        2\pi \hbar^2  
        \over  
                    m  
        }  
        n_o  
        \left[  
        b -i  
        {  
        b^2  
        \over  
           4\pi  
        }  
        {p_0\over \hbar}  
        \int d\Omega_q \,  
        S({\bf{q}})
        \right]  
        \]
which takes diffuse scattering into account, thus
straightforwardly recovering a result previously obtained
through multiple scattering theory. The incoherent
contribution instead accounts for fullfilment of the optical
theorem and is possibly responsible for loss of coherence in
interferometric experiments.
\par
The other application
concerns the so called quantum Brownian motion
(Vacchini, 2000). In
this case one considers the dissipative dynamics of a
particle interacting with a gas by two-body collisions.
Being interested in the local
dissipative behavior, we neglect the influence of the actual
boundary conditions, and suppose that the system may be
considered locally homogeneous within a very good
approximation, thus analyzing the interaction in momentum
space.
Under the assumption of small momentum transfers,
the balance between the anticommutator and
incoherent term leads to a quantum generalization of the
Fokker-Planck equation having a CP structure (Di\'osi, 1995),
while derivations starting at a fundamental level usually
missed positivity of the time evolution (Ambegaokar, 1991), let
alone CP.
The equation reads
        \begin{equation}
        \label{QBM}
        {  
        d {\hat \varrho}  
        \over  
                dt  
        }  
        =  
        -{i\over\hbar}  
        \left[  
        {{\hat {\mbox{\sf H}}}_0}
        + {{\hat {\mbox{\sf V}}}}
        ,{\hat \varrho}
        \right]  
        -
        {
        1
        \over
         \hbar^2
        }
        D_{pp}  
        \sum_{i=1}^3
        \left[  
        {\hat {{\sf x}}}_i,
        \left[  
        {\hat {{\sf x}}}_i,{\hat \varrho}
        \right]  
        \right]  
        -  
        D_{qq}
        \sum_{i=1}^3
        \left[  
        {\hat {\mbox{\sf p}}}_i,
        \left[  
        {\hat {\mbox{\sf p}}}_i,{\hat \varrho}
        \right]  
        \right]  
        -{i\over\hbar}
        D_{qp}
        \sum_{i=1}^3
        \left[  
        {\hat {{\sf x}}}_i ,
        \left \{  
        {\hat {\mbox{\sf p}}}_i,{\hat \varrho}
        \right \}  
        \right]
        \end{equation}
where ${{\hat {\mbox{\sf V}}}}$ is a mean-field potential,
$D_{pp}$ is expressed in terms of the scattering
cross-section and
        \[
        D_{qq}
        =
        \left(
        {
        1
        \over
             4MkT
        }
        \right)^2 D_{pp}    ,
        \qquad
        D_{qp}
        =
        {
        1
        \over
             2MkT
        }
        D_{pp}
        \]
$M$ being the mass of the particle. Equation (\ref{QBM})
actually has
a Lindblad structure (Barchielli, 1983) as it can be seen
by introducing the generators
$
        {{\hat {\mbox{\sf L}}}_i}
        =
        {\hat {{\sf x}}}_i
        +
        i
        (
        \hbar
        /
             2MkT
        )
        {\hat {\mbox{\sf p}}}_i
$,
thus obtaining
        \[
        {  
        d {\hat \varrho}  
        \over  
                dt  
        }  
        =  
        -{i\over\hbar}  
        \left[  
        {{\hat {\mbox{\sf H}}}_0}
        + {{\hat {\mbox{\sf V}}}}
        ,{\hat \varrho}
        \right]  
        -
        \frac{i}{\hbar}
        {
        D_{pp}  
        \over
             4MkT
        }
        \sum_{i=1}^3
        \left[  
        \left \{
        {\hat {{\sf x}}}_i,
        {\hat {\mbox{\sf p}}}_i
        \right \}
        ,{\hat \varrho}
        \right]  
        -  
        2
        {
        D_{pp}
        \over
             \hbar^2
        }
        \sum_{i=1}^3
        \left[  
        \frac 12
        \left \{
        {{\hat {\mbox{\sf L}}}_i^{\scriptscriptstyle\dagger}}
        {{\hat {\mbox{\sf L}}}_i}
        ,
        {\hat \varrho}
        \right \}
        -
        {{\hat {\mbox{\sf L}}}_i}
        {\hat \varrho}
        {{\hat {\mbox{\sf L}}}_i^{\scriptscriptstyle\dagger}}
        \right]  
        \]
\par
We will now briefly sketch how  the proposed formal scheme
may be applied to the simplest cases of systems having many
degrees of freedom (see Lanz {\it et al.},
1997; Lanz and Vacchini, 1998 for further details).
Slow variables inside a many-body  system,
characteristically
corresponding to densities of conserved charges, will have
the expression
        \[
        {\hat A}({{\mbox{\boldmath $\xi$}}})
        \sum_{hk}  
        { a}_h^{\scriptscriptstyle\dagger}
        A_{hk}({{\mbox{\boldmath $\xi$}}})  { a}_k
        \quad ,  
        \quad  
        {\hat B}({{\mbox{\boldmath $\xi$}}})
        \sum_{h_1 h_2 \atop k_1 k_2}
        { a}_{h_2}^{\scriptscriptstyle\dagger}
        { a}_{h_1}^{\scriptscriptstyle\dagger}
        B_{h_2 h_1 k_1 k_2} ({{\mbox{\boldmath $\xi$}}})
        { a}_{k_1}
        { a}_{k_2}
        \]
and similarly for quantities involving a higher number of
field  operators, where the couples of indexes $h_i, k_i$
are linked by a quasi-diagonality condition due to the slow
variability. One is therefore faced with evaluating in
Heisenberg picture on a time $t$ much longer than
collision times, but still short with respect to the
variation time of slow observables,
        \[
        {\cal U}'(t)({{a^{\scriptscriptstyle \dagger}_{h}}{a_{k}}})
        =
        e^{{i\over \hbar}{\hat  H}t}
        { a}_h^{\scriptscriptstyle\dagger}
        { a}_k
        e^{-{i\over \hbar}{\hat  H}t}
        \]
or more generally
        \[
        {\cal U}'(t)
        (
        a^{\scriptscriptstyle\dagger}_{h_n} \ldots
        a^{\scriptscriptstyle\dagger}_{h_1}
        a_{k_1} \ldots a_{k_n}
        )
        =
        e^{{i\over \hbar}{\hat  H}t}
        a^{\scriptscriptstyle\dagger}_{h_n} \ldots
        a^{\scriptscriptstyle\dagger}_{h_1}
        a_{k_1} \ldots a_{k_n}
        e^{-{i\over \hbar}{\hat  H}t}
        \]
Similarly as before, we ask that this map ${\cal U}'(t)$
satisfy a less stringent version of CP; in fact we ask CP
only when it is applied on these structures of blocks of
field  operators in the sufficiently diagonal case, i.e.,
when acting on the relevant, slowly varying observables.
Analogously as before calculations has been put forward
using a superoperator formalism, considering a
self-interacting Schr\"odinger field, i.e., a gas of
particles interacting through a short-range potential, and
working in a one-mode approximation, so  that three-particle
collisions are neglected. The result for the generator is
formally the same as before
        \[
        {\cal L}'
        \left(  
        {{{ a}^{\scriptscriptstyle \dagger}_{h}}{{ a}_{k}}}
        \right)  
        =
        {i\over\hbar}  
        \left[  
        {\hat H}_0 + {\hat V}^{[2]},
        { a}^{\scriptscriptstyle \dagger}_{h}
        { a}_{k}
        \right]  
        - {1\over \hbar}  
        \left\{
        \left[  
        {\hat \Gamma}^{[2]} , { a}^{\scriptscriptstyle\dagger}_h
        \right]  
        { a}_k
        -  
        { a}^{\scriptscriptstyle\dagger}_h
        \left[  
        {\hat \Gamma}^{[2]}, { a}_k
        \right]  
        \right\}
        +  
        {1\over\hbar} \sum_\lambda  
        {\hat R}^{[2]}_{h \lambda}{}^{\dagger}
        {\hat R}^{[2]}_{k \lambda}
        \]
but ${[2]}$ now denotes two-particle  operators and
statistical corrections are properly accounted for in the
structure of the T-matrix. A slight generalization of this
result holds in the case of $2n$ operators.
\par
We therefore hope to have shown new examples of useful
application of the property of CP, indicating how the
restriction of the requirement of CP to a suitable subset of
slowly varying quantities might be a guiding principle in
the determination of subdynamics inside nonrelativistic
quantum field theory.

\par
\vskip 15pt
\parindent = 0 pt
{\LARGE References}
\vskip 10pt

Alicki, R., and Lendi, K.
(1987).
Quantum Dynamical Semigroups and Applications, in {\it Lecture Notes in
Physics}, Vol. 286, {Springer}, {Berlin}.

Ambegaokar, V.
(1991).
{\it Berichte der Bunsengesellschaft f\"ur Physikalische
Chemie},
{\bf 95},
400.

Barchielli, A.
(1983).
{\it Nuovo Cimento},
{\bf 74B},
113.

Barchielli, A., Lanz, L., and Prosperi, G.~M.
(1983).
{\it Foundations of Physics},
{\bf 13},
779.

Birkhoff, G., and von Neumann, J. (1936).
{\it Annals of Mathematics},
{\bf 37},
823.

Busch, P., Grabowski, M., and Lahti, P. J.
(1995).
Operational Quantum Physics, in {\it Lecture Notes in
Physics}, Vol. m31, {Springer}, {Berlin}.

Dehmelt, H.
(1990).
{\it Review of Modern Physics},
{\bf 62},
525.

Di\'osi, L.
(1995).
{\it Europhysics Letters},
{\bf 30},  
{63}.

Dalla Chiara, M. L., Giuntini, R.
(n.d.).
Quantum Logic,
in {\it Handbook of Philosophical Logic},
Gabbay, D., Guenthner F., eds., Kluwer, Dordrecht, to appear

Hellwig, K.-E.
(1995).
{\it International Journal of Theoretical Physics},
{\bf 34},
{1467}.

Kraus, K.
(1971).
{\it Annals of Physics},
{\bf 64},
{311}.

Kraus, K.
(1983).
States, Effects and Operations, in {\it Lecture Notes in
Physics}, Vol. 190, {Springer}, {Berlin}.

{Lanz, L., Melsheimer, O., and Vacchini, B.}
(1997).
Subdynamics through time scales and scattering maps
in quantum field theory,
in  
{\it Quantum communication, computing, and measurement},
Hirota, O., Holevo, A.~S., and Caves, C.~M. eds., Plenum, New York,
pp. 339-353.

{Lanz, L., and Vacchini, B.}
(1997a).
{\it International Journal of Theoretical Physics},
{\bf 36},
{67}.

{Lanz, L., and Vacchini, B.}
(1997b).
{\it Physical Review A},
{\bf 56},
{4826}.

{Lanz, L., and Vacchini, B.}
(1998).
{\it International Journal of Theoretical Physics},
{\bf 37},
{545}.

{Lindblad, G.}
(1976).
{\it Communications in Mathematical Physics},
{\bf 48},
{119}.

{Ludwig, G.}
(1983).
{\it Foundations of Quantum Mechanics}, {Springer}, {Berlin}.

Schumacher, B.
(1996).
{\it Physical Review A},
{\bf 54},
{2614}.

{Srinivas, M.~D., and Davies, E.~B.}
(1981).
{\it Optica Acta},
{\bf 28},
{981}.

Streater, R.~F.
(1995).
{\it Statistical Dynamics},
Imperial College Press, London.

{Vacchini, B.}
(2000).
{\it Physical Review Letters},
{\bf 84},
1374.
\end{document}